\documentstyle[aps,amssymb]{revtex}

\input epsf

\begin{document}
\title{Phase Transitions in a Driven Lattice Gas with Anisotropic Interactions%
\medskip }
\author{L. B. Shaw, B. Schmittmann and R. K. P. Zia\bigskip}
\address{Center for Stochastic Processes in Science and Engineering\\
Department of Physics \\
Virginia Polytechnic Institute and State University\\
Blacksburg, Virginia 24061-0435, USA}
\date{\today}
\maketitle

\begin{abstract}
The Ising lattice gas, with its well known equilibrium properties, displays
a number of surprising phenomena when driven into non-equilibrium steady
states. We study such a model with anisotropic interparticle interactions ( $%
J_{\Vert }\neq J_{\bot }$ ), using both Monte Carlo simulations and high
temperature series techniques. Under saturation drive, the shift in the
transition temperature can be both positive and negative, depending on the
ratio $J_{\Vert }/J_{\bot }$! For finite drives, both first
and second order transitions are observed. Some aspects of the phase diagram
can be predicted by investigating the two point correlation function at the
first non-trivial order of a high temperature series expansion.
\end{abstract}

\bigskip
\section{Introduction}

Our understanding of collective behavior in many-body systems in, or near,
thermal equilibrium has improved significantly in the past century. By
contrast, only recently has there been much attention focused on systems
far-from-equilibrium, even for those in steady states. Abundant in nature,
such systems exist in an enormous variety of states, most of which cannot be
predicted using the framework of equilibrium statistical mechanics. At
present, there is no simple formulation which is generally applicable. One
way to make progress into this vast realm of non-equilibrium physics is to
study simple models, especially those with well understood equilibrium
properties. This is the chief motivation of Katz, et.al.,\cite{kls} who
introduced a seemingly trivial modification to the Ising lattice gas\cite
{ising}, so that it is driven into {\em non-equilibrium steady states}. The
other motivation is phase transitions in fast ionic conductors in an
external field, though their model predictions have not been compared
directly with these physical systems so far.

This model consists of a $d$-dimensional hyper-cubic lattice with each site
being empty or occupied by a single particle. Its dynamics is particle
hopping to nearest neighbor empty sites, with a rate controlled by the
energetics of the Ising Hamiltonian ${\cal H}$, as well as a bias in one
direction so as to describe the effect of a uniform, DC ``electric'' field $%
E $. To connect with the equilibrium system, rates are chosen to satisfy
``local'' detailed balance consistent with being in contact with a thermal
bath at temperature $T$. In particular, by setting $E=0$, we simply retrieve
the canonical $\exp \left( -{\cal H}/k_BT\right) $ as the stationary
distribution. Since its inception, many unexpected properties have been
discovered, for both the original model and its many variants \cite{rev}.
Though some of the remarkable behaviors are reasonably well understood,
many remain unexplained. An example is the basic question: why should $%
T_c(E) $, the critical temperature for the attractive (ferromagnetic) case,
increase with $E$? Simulations (using Metropolis rates for $d=2$) show that $%
T_c(\infty )$ is about 40\% above the Onsager temperature \cite{tc}! Indeed,
one can easily argue in favor of a {\em lowering} of $T_c$, as follows.
Since large fields would dominate over the inter-particle attraction (for
hops along $E$), the system is effectively subjected to an extra noise, so
that a lower $T$ is needed for clustering and order.

With further explorations of this system, simple arguments in favor of an 
{\em increased} $T_c$ began to emerge\cite{bilayer}. However, there is still
no {\em compelling} reason, so far, to accept or reject the contradictory
arguments. In other words, we still have no intuitive picture which can
guide us to reliable expectations. This study is a continuation of the
exploration of similar systems \cite{bilayer,SPBCOBC,am}, in an effort to
find the underlying mechanisms which give rise to the novel phase diagrams.
Specifically, we consider Ising models with {\em anisotropic} interactions,
i.e., inter-particle attraction along and transverse to the drive being
unequal ($J_{\parallel }$ $\neq $ $J_{\perp }$). Such a model was previously
studied, but only in the fast rate limit \cite{FRL}. Unfortunately, though
very interesting in its own right, this limit is too singular to provide us
with much insight into the original lattice gas. For example, no trace of $E$
remains, yet a phase transition prevails even for $J_{\perp }\equiv 0$! In
the following section, we describe our model, restricting our attention to $%
d=2$ only. Simulations lead to more surprises, the details of which will be
presented in Section 3. The next section is devoted to a theoretical study,
using high temperature series expansion techniques on the two-point
correlation function \cite{ZWLV,HTS}. A final section is devoted to a
summary, conclusions and outlook. 
\[
\]

\section{Driven lattice gas with anisotropic interactions}

Our system consists of fully periodic $L\times M$ square lattices, with the
sites labeled by ${\bf x}=(x,y)$, where $x$ and $y$ are integers modulo $L $
and $M$, respectively. Each site may be empty or occupied by a particle, so
that a configuration of the system is specified by the set of occupation
numbers $\{n({\bf x})\}$, where $n=0\ $or $1$. This model is simply related
to the original Ising one for spins through the relation $s\equiv 2n-1$,
which takes on values $\pm 1$. Since our interest is a system of particles, $%
\sum_{{\bf x}}n({\bf x})$ is fixed. For simplicity, we study only
half-filled systems, i.e., $\sum n=LM/2$ (or $\sum s=0$). Next, we endow the
particles with {\em anisotropic }nearest neighbor attraction and write the
Hamiltonian as 
\begin{equation}
{\cal H}\equiv -4J_x\sum n(x,y)n(x+1,y)-4J_y\sum n(x,y)n(x,y+1)  \label{Ham}
\end{equation}
with $J_x,J_y>0$. The factor of 4 is included so that the Hamiltonian is
equivalent (given the constraint $\sum s=0$) to the standard Ising form, $%
-J\sum ss^{\prime }$. For convenience, we introduce anisotropy via a
``dimensionless'' parameter $\alpha $, through 
\begin{equation}
J_x=J/\alpha \quad \text{and\quad }J_y=J\alpha  \label{alpha}
\end{equation}
so that $J$ carries the ``overall'' strength of the interactions. When this
system reaches equilibrium with a thermal bath at temperature $T$, it
displays well known properties in the thermodynamic limit
\cite{Onsager,LY,McWu}. 
The most prominent behavior is a second order transition, from a
disordered homogeneous phase to a phase segregated state, when $T$ is
lowered through the Onsager temperature $T_c(\alpha )$. Expressed in units
of $J/k_B,$ where $k_B$ is Boltzmann's constant, $T_c(\alpha )$ can be
obtained from the equation \cite{K-W} 
\begin{equation}
\left( 1+\varepsilon ^\alpha \right) \left( 1+\varepsilon ^{1/\alpha
}\right) =2  \label{Tc(alpha)}
\end{equation}
for the quantity $\varepsilon \equiv \exp \left( -2J/k_BT_c(\alpha )\right) $%
. Specifically, $T_c(1)\simeq 2.269J/k_B$, which serves as the unit for all
the temperatures quoted in this paper. Using this unit, a plot of $%
T_c(\alpha )$ is provided in Fig. 1. Note that, due to the conservation law,
the ordered state will be a strip spanning the system aligned with the $x$-
or $y$-axis. These states will be denotes by H (horizontal) and V
(vertical), respectively. Since the equilibrium state will be the one with
the lowest interfacial free energy, the aspect ratio of the system, $L/M$,
will play a role as well. For $L/M=1$, clearly $\alpha =1$ marks the ``phase
boundary'' between the two states, also shown in Fig. 2. We expect, of
course, that the H-V boundary is associated with a first order transition.
For general $L/M$, this boundary is located at that $\alpha (T)$ which
satisfies $\sigma _y(\alpha ,T )/\sigma _x(\alpha ,T)=L/M$, where $\sigma
_x$ is the surface tension for the horizontal
interface, etc.

In computer studies, the simulation of the lattice gas, in equilibrium
with a bath at temperature $T$, would rely on ``spin-exchange'' dynamics\cite
{kawa}, so as to respect particle conservation. In other words, particles
are allowed to hop to nearest neighbor holes with a rate obeying detailed
balance appropriate for $T$. A favorite rate is due to Metropolis \cite
{metro}, where jump rates are given by $\min \{1,\exp {(-\Delta {\cal H}%
/k_BT)}\}$, where $\Delta {\cal H}$ is the change in energy due to the hop.
In the generalization to a {\em driven} lattice gas, we follow Katz, et. al.
and incorporate the ``electric'' field (aligned with the $y$-axis) in the
standard way, i.e., by adding $\pm E$ to $\Delta {\cal H}$ for hops
against/along the field \cite{kls,rev}. Our goal is to map out the phase
diagram in the $T$-$\alpha $-$E$ space. In the following, we will report the
main results, denoting the line of second (first) order transitions by $%
T_c(\alpha ,E)$ ($\alpha _1(T,E)$) and quoting $E$ in units of $J$. Details
will be published elsewhere \cite{big}. 

\section{Simulation methods and results}

Most of our runs involve lattices with $L=M=30$. For definiteness, we
aligned our drive so that particles are biased to hop ``downwards'' ($-y$
direction). Starting with one of three initial configurations of a
half-filled lattice, the system is evolved in the standard way. In a Monte
Carlo step (MCS), $2LM$ bonds (nearest neighbor pairs) are chosen at random.
If the bond consists of a particle hole pair, then the pair is exchanged
with probability $\min \{1,\exp {(-}\left[ {\Delta \mathcal{H}}+E\delta
y\right] {/k_BT)}\}$, where $\delta y$ is the change in the $y$-coordinate
(mod $M$) of the \emph{particle}. Most runs are 400K MCS long. The initial
configurations are either random or fully ordered (in H or V). Typically, we
discard the first 100K MCS, allowing the system to settle into the steady
state. Then, every 200 MCS, we measure the Fourier transform of the particle
density $n(\mathbf{x})$%
\begin{equation}
\tilde{n}(k,p)\equiv \frac 1{\sqrt{LM}}\sum_{x,y}n(x,y)\ \exp \left[ 2\pi
i\left( \frac{kx}L+\frac{py}M\right) \right]  \label{FT}
\end{equation}
where $k,p$ are integers. Averaging over the rest of the run (denoted by $%
\langle \rangle $), we obtain the structure factors 
\begin{equation}
S(k,p)\equiv \langle |\tilde{n}(k,p)|^2\rangle .  \label{S}
\end{equation}
In the disordered state, they are of order unity and convey information
about two-particle correlations. Below the transition, they are sensitive to
the H- or V-ordered state, in the sense that, e.g., $S(0,1)$ or $S(1,0)$
will be $O(LM)$, and carry information about the densities of phase
segregated states. For example, for a completely ordered V state, we obtain $%
S(0,1)\equiv 0$ while $S(1,0)=\left( 2M/L\right) \left[ 1-\cos (2\pi
/L)\right] ^{-1},$ which approaches $LM/\pi ^2$ in the thermodynamic limit.
Of course, in an ordered H state, these two $S$'s are reversed. In this
sense, $S(0,1)$ and $S(1,0)$ may be viewed as order parameters\cite{kls}. In
addition to the mean of $|\tilde{n}|^2$, we measure the variance, $\left(
\Delta S\right) ^2\equiv \langle |\tilde{n}|^4\rangle -S^2$, which plays a
role similar to that of the susceptibility in equilibrium systems. For those 
$S$'s which are $O(1),$ the associated $\Delta S$'s are expected to be just $%
S$ itself \cite{Rudzin}. Deep in the ordered phases, though some $S$'s will
be $O(LM)$, the $\Delta S$'s are expected to be still $O(1)$. However, near
a second order transition, the $\Delta S$'s should diverge in the
thermodynamic limit. We have also exploited other initial configurations,
different sizes and rectangular systems. These will be discussed in the
following sub-sections.

\subsection{Second order transitions with saturation drive}

In the first part of our study, we restrict ourselves to saturation drive (
$E=50 $ here) and to only one system size/geometry: $30\times 30$. Since the
field is large enough to overcome all levels of particle attraction, it is
hardly surprising that the ordered state can only be V. Measuring the
structure factors, we confirm this expectation, in that only $S(k,0)$ ($k$
odd) grow substantially as $T$ is lowered. Specifically, starting with
random configurations, we perform 400 K\ MCS runs for $\alpha =\frac
13,\frac 12,\frac 34,1,\frac 43,2,3$ and $0.5\leq T\leq 3.0$. We used a
range of step sizes in $T$, the smallest being $0.025$, especially when
large variations in measured quantities were encountered. Identifying the
transition through the peak of $\left( \Delta S(1,0)\right) ^2,$ we find the
line of second order transitions $T_c(\alpha ,\infty )$ to be a \emph{%
monotonically decreasing} function of $\alpha $, as shown in Fig. 1. This
behavior is contrary to our previous conjecture\cite{bilayer}, showing that
our intuitive picture, based on the competition of long and short ranged
correlations, is still far from being reliable. More surprisingly, the
critical temperature of the driven system is not always greater than that
for the equilibrium system! Note that, in Fig. 1., $T_c(\alpha ,0)$ is the
theoretical result for an infinite lattice, while $T_c(\alpha ,\infty )$ is
found in the $30\times 30$ system. (The effects of finite size will be
discussed in a later sub-section.) A further distinction is that, while the
driven system orders only into the V state, the equilibrium system may order
into H or V depending on $\alpha <1$ or $>1$.

We should caution that the uncertainties associated with $T_{c}(\alpha
,\infty )$ are much larger than the step size of $0.025.$ In particular, for 
$T$ greater than the quoted $T_{c}(\alpha ,\infty )$, $S(1,0)$ does not
decrease in a simple way, especially for systems with large $\alpha .$ In
these cases, we frequently observe breakup of a single strip into two-strip
states. As a result, as $T$ is raised through this $T_{c}(\alpha ,\infty )$
, not only does $\left( \Delta S(1,0)\right) ^{2}$ becomes large, $S(2,0)$
also increases by as much as an order of magnitude. In this sense, it is
possible to regard the quoted values as a lower bound for the critical
temperature. Typically, at temperatures $\thicksim $ 30\% higher than $%
T_{c}(\alpha ,\infty )$, we can be quite certain that the system is
disordered. To be more confident and precise about critical temperatures, we
must use more sophisticated methods, e.g., histograms of various $S$'s,
finite size scaling, etc.\cite{big}

\subsection{First and second order transitions with small drives}

Next, we turn to systems with small drives, searching for the remnants of
the H phase. Keeping the same system size ($30\times 30$), we chose $E=1,2,$
and $5$, so that there is a range of values for the ratio $E/J_{\bot
}=\alpha E$. Expecting continuous transitions in the neighborhood of the
equilibrium line, we performed simulations in the manner described above. As
displayed in Fig. 2, the behavior of $T_c(\alpha ,E)$ is quite complex. For $%
\alpha =3$, $T_c$ first increases and then decreases, so as to match with
the low value of $T_c(3,\infty )$ found above. For $\alpha =1$, the known
monotonic increase of $T_c$ is confirmed. In its neighborhood, both $%
T_c(\frac 43,E)$ and $T_c(\frac 34,E)$ appear to be monotonic in $E$. In the
latter case, the ordered phase remains V, despite the fact $\alpha <1$. This
reflects the known effect of the drive: setting up positive long range
correlations along the drive and so, favoring strips aligned vertically. For 
$\alpha <3/4$ and the $E$'s we have chosen, the systems orders into V or H
depending on the drive. For those cases where the ordered phase is H, it is
completely unexpected that $T_c(\alpha ,E)$ is still typically \emph{greater}
than the equilibrium $T_c(\alpha ,0)$! Since the drive is known to \emph{%
reduce} correlations in the transverse direction, we must find a new
argument for how ordering (into H) could occur at a higher temperature.
Remarkably, $T_c(\alpha ,E)$ suffers a dip at the bi-critical point. As a
result, for a fixed $\alpha $, it is seen first to increase with $E$ (while
ordering into H), then to decrease to the minimum value $T_c(\alpha ,E_B)$
when the drive reaches $E_B$, a strength just strong enough to change the
ordering to V, and finally, to increase again with stronger drives. Given
our accuracy, $T_c(\alpha ,E_B)$ seems to be the same as $T_c(\alpha ,0),$
though we have no reason to believe that they are indeed equal!

\begin{figure}[tbph]
\hspace*{1.1cm} \epsfxsize=6in \epsfbox{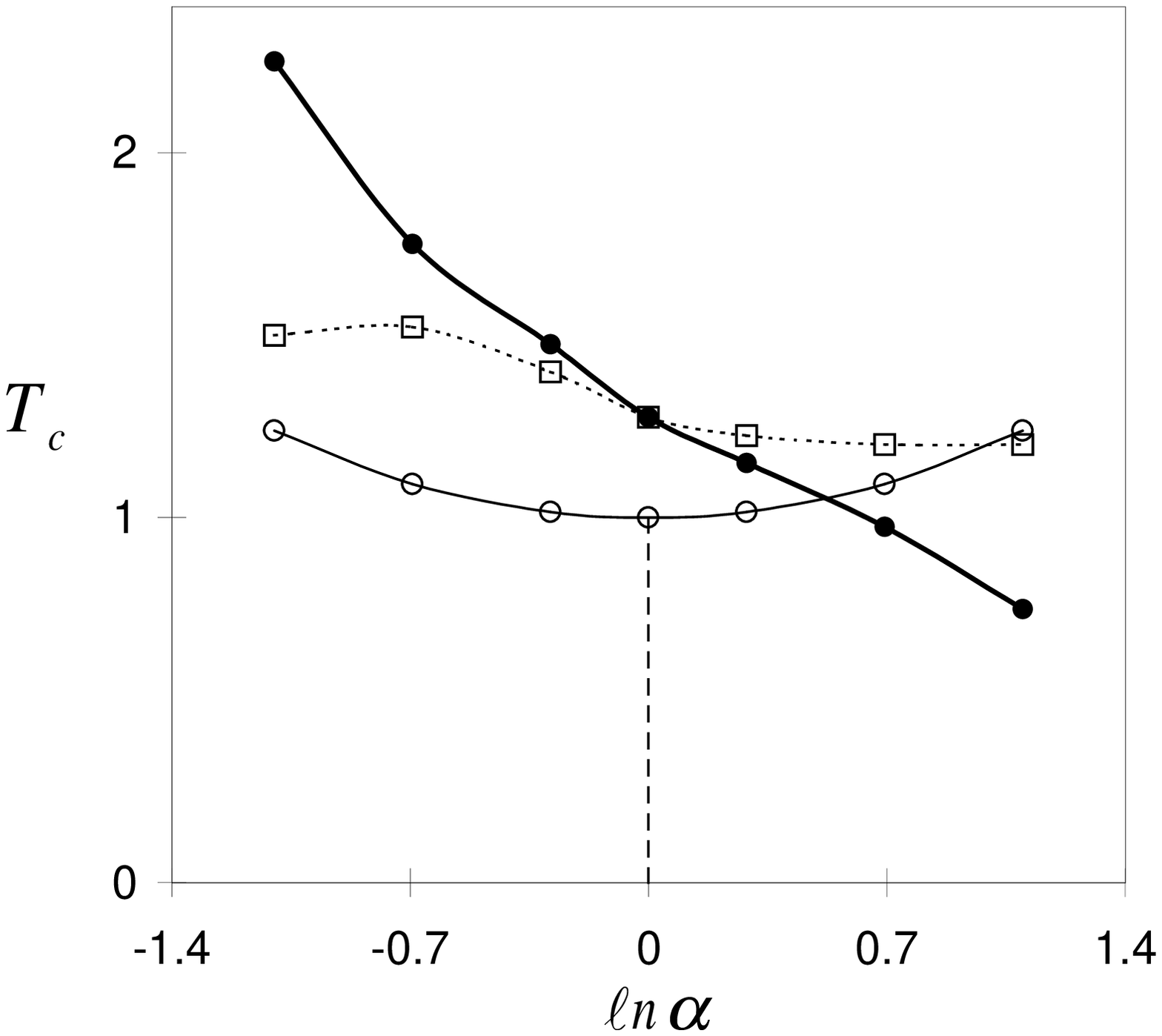}
\vspace*{-1.7cm}
\end{figure}
\begin{center}
Figure 1. $\quad$ Critical temperatures vs. $\ell n\alpha $ for
$E=5(\square )$ and $50(\bullet )$. Both ordered states are V.

The exact values for $E=0$ are labelled by $\circ $'s;
the dashed line is the boundary between H- and V-states.
\end{center}
\vspace*{-.4cm}
\begin{figure}[htbp]
\hspace*{1cm}
\epsfxsize=6in \epsfbox{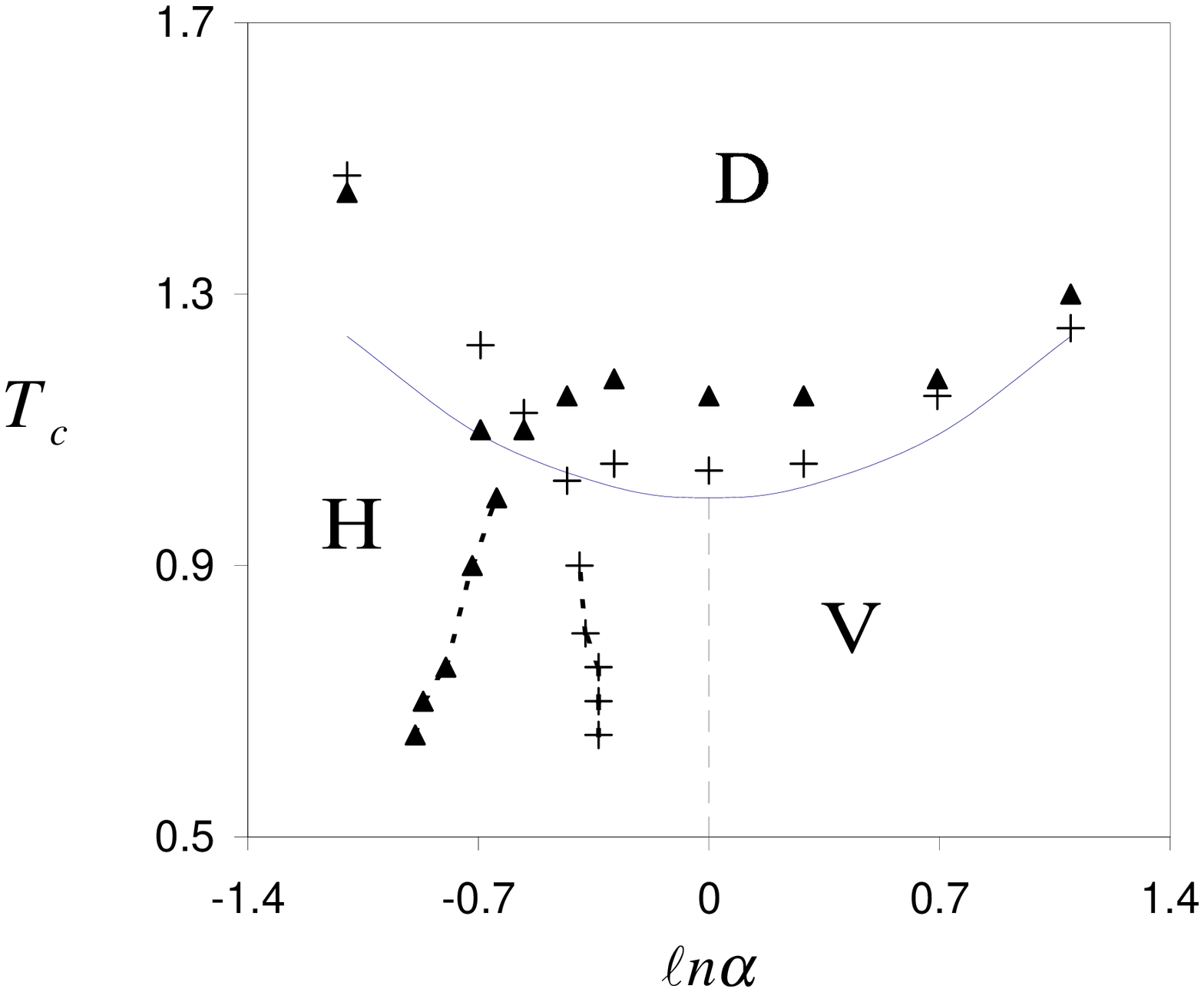} 
\vspace*{-1.7cm}
\end{figure}
\begin{center}
Figure 2. $\quad$ Phase Diagram for $E=0$ (solid and dashed lines), $1 $ (+),
and $2$ ($\blacktriangle $).

Points associated with first order transitions are joined by a dotted line.

The disordered region is labelled by D; the ordered ones by H and V (See
text).
\end{center}
\vspace*{.5cm}

Turning to the line of discontinuous transitions, we modified our methods.
Since it is expected to be more or less aligned with the $T$ axis, we sweep
in $\alpha $ with \emph{fixed} $T$ and $E$. The critical values are denoted
by $\alpha _1(T,E),$ shown as points joined by dotted lines in Fig.2.

For $E=1,2$ and $T$ as low as $0.9$, we are able to look for hysteresis in
such sweeps. In both cases, $\alpha $ is increased and decreased by steps of 
$0.025$ and the system evolved for 200K MCS at each step. The ratio 
\[
\frac{S(0,1)-S(1,0)}{S(0,1)+S(1,0)} 
\]
is monitored and is seen to jump from 1 to -1 and vice versa. $\alpha
_1(T,E) $ is found by averaging the values of $\alpha $ where the jump
occurs.

For lower $T$, the metastable life times are too long and hysteresis loops
become too large to be reliable. Thus, we resort to another method, which,
to our knowledge, is entirely new. Though it does not \emph{necessarily}
identify the loci of first order transitions, we argue that it should
provide a good indication. This approach relies on a conjecture of the
saddle point configuration, namely, an ``X'' (Fig. 3a), which is chosen via
symmetry considerations. Given the anisotropies in the system (due to both $%
\alpha $ and $E$), we concede that the saddle point may be a configuration
with less symmetry, e.g., an ``L'' (Fig. 3b). Nevertheless, we believe that
the ``X'' should be a good starting point. First, we carry out 400K MCS runs
with V and/or H as the initial configuration in this region of phase space.
We chose only three values of $T:0.65,0.70,$and $0.75$. For $E=1$, we
focused on the values $\alpha =0.65,0.70,0.75,$ and $0.80$. For $E=2$, we
used $\alpha =0.40,0.45,0.50,$ and $0.55$. Since the system remains in the
initial state, sharp distributions of the structure factors ($%
S(1,0),S(2,0),S(0,1),$ and $S(0,2)$) emerge, providing us with good averages
and standard deviations. Next, we carried out 200 independent runs for each
of the above ($T,\alpha ,E$), \emph{starting with the }``X''\emph{\
configuration}. The above $S$'s are measured every 200 MCS. The runs are
terminated when 10 consecutive measurements (i.e., over a period of 2 K MCS)
of each $S$ fall within 3 standard deviations of the averages. The fraction
of runs which terminate in the V state is recorded and found to increase
monotonically with $\alpha $ (with fixed $T,E$). The points where this
fraction reaches 1/2 are considered (part of) the line of first order
transitions from H to V. (See Fig. 2.) 
\vspace*{.5cm}
\begin{figure}[htbp]
\hspace*{3.5cm}
\epsfxsize=4in \epsfbox{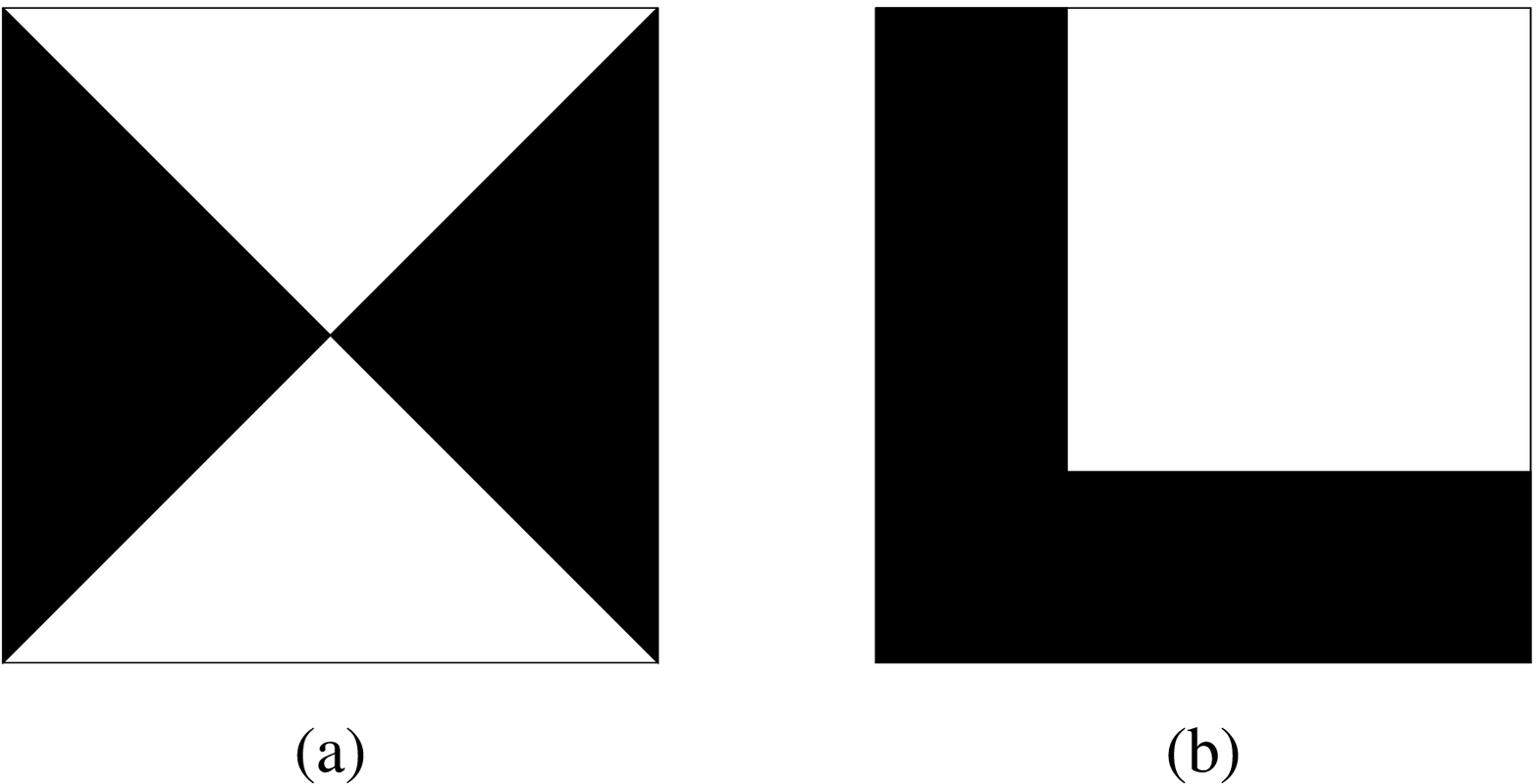} 
\end{figure}
\begin{center}
Figure 3. $\quad$ Possible saddle point configurations: (a) ``X'' and
(b) ``L'.
\end{center}
\vspace*{.5cm}

Finally, for $E=5$, all ordered states we have observed are V. Thus, we
never explored the line of first order transitions. To keep Fig.2 relatively
clear, we display this line of continuous transitions in Fig.1 instead. It
is possible that first order transitions occur just outside the range of $%
\alpha $'s we used. Indeed, the slight dip of $T_c(\alpha ,5)$ at $\alpha
=1/3$ is actually due to behavior which is more complicated than that in
systems with higher $\alpha $. For $T\geq 1.6$, $S(1,0)<S(0,1)$, even though
the system actually orders into V for $T\leq 1.5$, i.e., $S(1,0)\gg S(0,1).$
In between, the fluctuations of both structure factors are comparably large.
In this sense, it is possible that ($\alpha =1/3,T\sim 1.55$) is very close
to the bi-critical point. To explore further, we must set $\alpha <1/3$.
However, we believe that, with $J_{\bot }/J_{\Vert }\gtrsim 10$, reliable
conclusions cannot be drawn from simulations with a 
lattice as small as $30\times 30$ .

\subsection{Effects of varying system sizes and aspect ratios}

The above data were collected on a single system size: $30\times 30$. Of
course, true phase transitions accompanied by thermodynamic singularities
are properties of infinite systems. Any conclusions would rely on either
simple extrapolations or more sophisticated methods of finite size scaling.
With our limitations of computation power, we are able to make only cursory
explorations of the effects of finite size. In particular, we studied $%
60\times 60$ and $90\times 90$ systems, only for $E=50$ and the extreme $%
\alpha $'s. The transition temperatures are summarized in Table 1. Similar
to the equilibrium cases, the transition temperatures tend to rise with
larger sizes. Although there are some significant increases, we can safely
conclude, from the trends displayed here, that the general features of
Figs.1 and 2 will survive the thermodynamic limit.
\begin{center}
Table 1. \quad Transition temperatures for various $L$ and 
$\alpha .$ \bigskip

\begin{tabular}{|r|c|c|c|c|}
\hline
$\ \quad L\diagdown \alpha $ & 1/3 & 1/2 & 2 & 3 \\ \hline\hline
\multicolumn{1}{|c|}{30} & \multicolumn{1}{||c|}{2.25} & 1.75 & 0.97 & 0.67
\\ \hline
\multicolumn{1}{|c|}{60} & \multicolumn{1}{||c|}{2.40} & 1.85 & 1.05 & 0.77
\\ \hline
\multicolumn{1}{|c|}{90} & \multicolumn{1}{||c|}{2.45} & 1.90 & 1.05 & 0.80 
\\ \hline
\end{tabular}
\end{center}

Finally, we also investigated the effects of having rectangular systems. By
varying the aspect ratio, we can insure that the \emph{ground state} of the
equilibrium system is in the V- or the H-state, with the crossover point at $%
M/L=\alpha ^2$. We could study systems with these aspect ratios. On the
other hand, at any finite temperature, the ``crossover'' $M/L$ is given by
the ratio of surface energies associated with the interfaces \cite{sigmas}
aligned along the two axes, i.e., 
\begin{equation}
\frac ML=\frac{2K\alpha +\ell n\left( \tanh \left( K/\alpha \right) \right) 
}{2K/\alpha +\ell n\left( \tanh \left( K\alpha \right) \right) }\quad ;\quad
K\equiv J/k_BT\quad .  \label{M/L}
\end{equation}

In other words, for a given aspect ratio, this equation provides the phase
boundary between H- and V-states. By duality, this ratio is also the one for
the correlation lengths associated with two-point correlations along the two
axes\cite{dual} in the disordered phase. Thus, another possibility is to
study systems which, near criticality, behave isotropically when the
rectangular systems are rescaled to be square ones. Since our interest is
mainly in disorder-order transitions, we adopted the latter choice here.
Evaluating (\ref{M/L}) near $T_c$, we arrived at lattice sizes $25\times 36,$
$20\times 48,$ and $15\times 64$ (for $\alpha =4/3,$ $2,$ and $3$
respectively), in order to compare with the $30\times 30$ case at $\alpha =1$%
. By symmetry, systems with the opposite aspect ratio are used for $\alpha
<1 $. Keeping in mind the previous discussion concerning the uncertainties
associated with $T_c$, we display the results in Fig.4. Note that the
equilibrium line is \emph{not shown}, since it is so close to the $E=1$
points that confusion may arise. It is unclear if this insensitivity of $%
T_c(\alpha ,1)$ to the drive is significant or not. Note that for $\alpha
=1/3$ and $1/2$, the system ordered into an H-state. For $E=2$, the lowering
of $T_c(1/3,2)$ is curious. This point also coincides with ordering into an
H-state. Comparing with Fig. 2, we see that there are two main effects due
to rectangular lattices: (i) displacement of the first order line to smaller 
$\alpha $'s and (ii) lowering of the transition temperatures into the
H-state. The lack of a dip in the $E=5$ case is consistent with the
bi-critical point being at $\alpha \lesssim 1/3$. Apart from these features,
system geometry seems to play no discernible role, especially for large
drives. That the system subjected to small $E$ orders into H-states may be
related to the fact that, at low temperatures, the H-state is favored in the
equilibrium systems according to eqn.(\ref{M/L}). Clearly,
there remains a large gap between our understanding of these systems and the
rich phenomena displayed.

\section{Two point correlation in a high temperature expansion}

Though the driven lattice gas model was introduced 15 years ago, there is
still no {\em reliable} means to predict the general features of the phase
diagram. The dynamic mean-field approach \cite{DMF}, while quantitatively
more satisfying, is so labor-intensive that it provides us with little
insight on why critical temperatures shifts to higher values. This method
can certainly be extended to our anisotropic system. Another route to an
estimate of the shift of $T_c$ is based on a recent study of the two-point
correlation function $G({\bf x})$ in a ``high temperature'' expansion\cite
{HTS}, which is actually an expansion in small $J$ or $K=J/k_BT$. In this
approach, an approximate equation for $G({\bf x})$, first derived in \cite
{ZWLV} is solved exactly. The resultant Fourier transform is just the
theoretical structure factor: $S({\bf k})$ which not only displays the well
understood discontinuity singularity at ${\bf k}=0$ \cite{ZHSL}, but also
may be exploited to estimate $T_c(E)$ \cite{HTS}. The results are less
accurate than those from dynamic mean-field theory, since the latter is the
generalization of the Bethe-Peierls approximation for equilibrium Ising
models. However, its implementation is much simpler. Here, we provide a
brief presentation of the generalization to the anisotropic model.

\begin{figure}[tbph]
\hspace*{1cm} \epsfxsize=6in \epsfbox{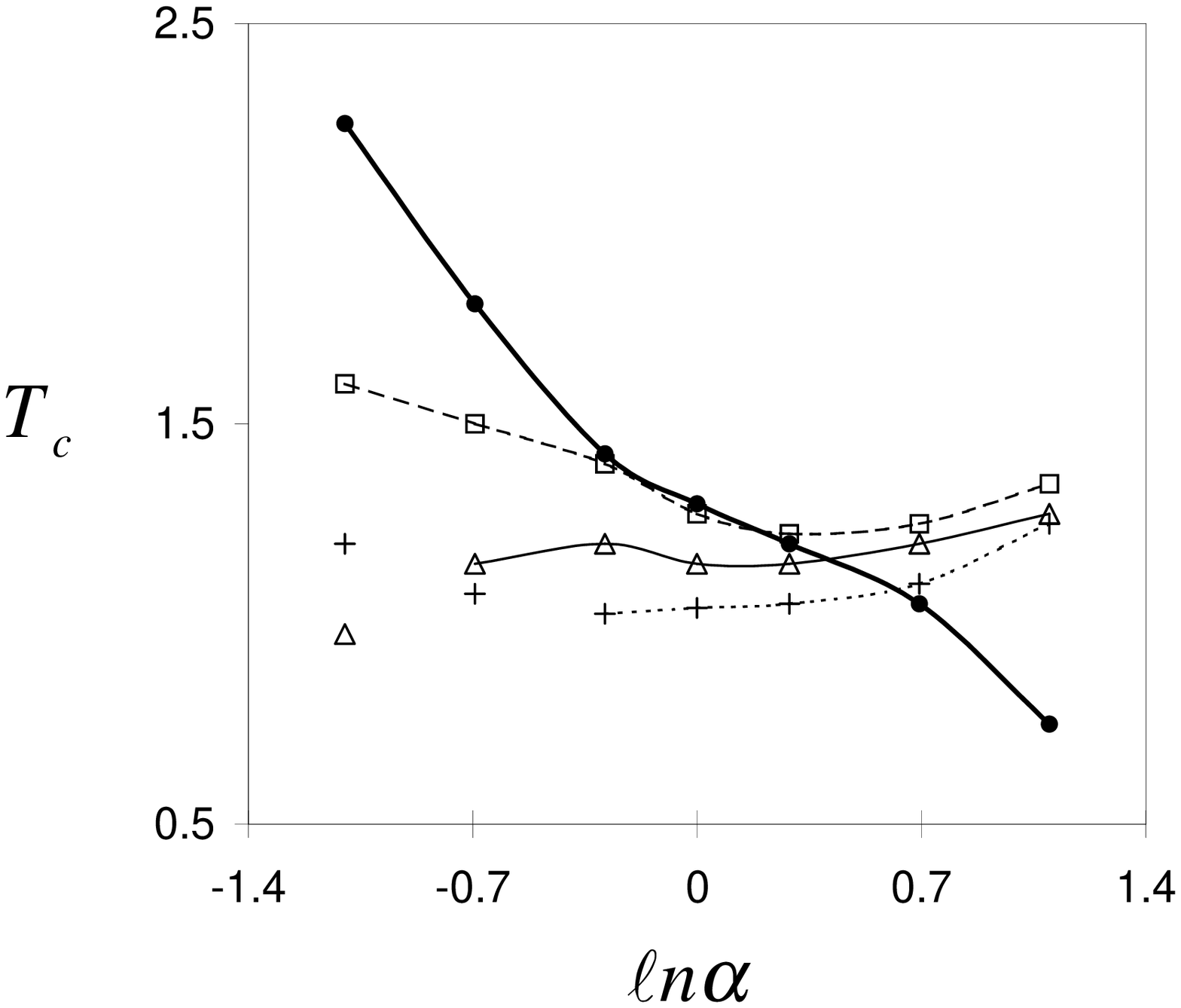}
\vspace*{-1.5cm}
\end{figure}
\begin{center}
Figure 4.\quad Transition temperatures for rectangular systems as
a function of $\ell n\alpha $.

Different drives are shown as $E=1(+),$ $2(\vartriangle ),$ $5(\square ),$
and $50(\bullet )$.

Transitions into the V-state are joined by lines.
\end{center}
\vspace*{.5cm}

With reference to \cite{ZWLV,HTS}, where the derivation and the nature of
the approximation can be found, we simply quote the equation for $G:$ 
\begin{eqnarray}
\partial _tG(0,1) &=&2[G(1,1)+G(-1,1)-2G(0,1)]
+(1+\epsilon )[G(0,2)-G(0,1)]+2K(2+\epsilon )\alpha \nonumber \\
\partial _tG(1,0) &=&2[G(2,0)-G(1,0)] 
+(1+\epsilon )[G(1,1)+G(1,-1)-2G(1,0)]+2K(1+2\epsilon )/\alpha  
\nonumber \\
\partial _tG(1,1) &=&2[G(2,1)+G(0,1)-2G(1,1)]  \label{Gnr0}
+(1+\epsilon )[G(1,2)+G(1,0)-2G(1,1)]-2K(\alpha +\epsilon /\alpha ) \\
\partial _tG(0,2) &=&2[G(1,2)+G(-1,2)-2G(0,2)] 
+(1+\epsilon )[G(0,3)+G(0,1)-2G(0,2)]-2\epsilon K\alpha  \nonumber \\
\partial _tG(2,0) &=&2[G(3,0)+G(1,0)-2G(2,0)]
+(1+\epsilon )[G(2,1)+G(2,-1)-2G(2,0)]-2K/\alpha  \nonumber
\end{eqnarray}
and, for all other non-zero $x,y$: 
\begin{eqnarray}
\partial _tG(x,y) &=&(1+\epsilon )[G(x,y+1)+G(x,y-1)-2G(x,y)]
+2[G(x+1,y)+G(x-1,y)-2G(x,y)]  \label{Gfar}
\end{eqnarray}
Here, 
\[
\epsilon \equiv \exp (-E/k_BT) 
\]
with the understanding that these equations are {\em valid to first order}
in $K$. Setting the left hand side of (\ref{Gnr0}) to zero, the linear
equations for the steady state $G^{*}(x,y)$ and its transform $S^{*}(k,p)$
can be solved, with the proviso that, at the zeroth order in $K,$ 
\begin{eqnarray*}
G^{*}(0,0) &=&1 \\
G^{*}(x,y) &=&0\ \ \text{ }x,y\neq 0\text{,}
\end{eqnarray*}
and 
\[
S^{*}(k,p)=1. 
\]

Since $\alpha $ does not affect the structure of these equations and enters
only through the inhomogeneous terms, we can follow the analysis in \cite
{HTS} closely and obtain the equation for $T_c(\alpha ,E)$. The result can
be succinctly summarized by 
\begin{equation}
\frac{T_c(\alpha ,E)}{T_c(1,E)}=\frac 1{1+f(\epsilon )}\alpha +\frac{%
f(\epsilon )}{1+f(\epsilon )}\frac 1\alpha  \label{TcRatio}
\end{equation}
The structure of this ratio is necessarily of this form, since $T_c(\alpha
,E)$ is linear in ($J_{\Vert },J_{\bot }$)=$J(\alpha ,1/\alpha )$ and, of
course, the ratio is unity when $\alpha =1.$ Also, since this approach
reduces to the simple mean-field result\cite{MFTc}, $\alpha +1/\alpha ,$ for
the equilibrium case ($E=0$), we know $f(1)=1$. Unfortunately, the
expression for $f$ 
\begin{equation}
f(\epsilon )=\frac{R_{11}+2(1-\epsilon )\left[ R_{10}R_{12}-R_{11}^2\right] 
}{R_{11}+2(1-\epsilon )\left[ R_{10}R_{21}-R_{20}R_{11}\right] }\quad ,
\label{f}
\end{equation}
where 
\begin{equation}
R_{ij}\equiv \frac 1{(2\pi )^2}\int_{-\pi }^{+\pi }d\tilde{k}d\tilde{p}\frac{%
(1-\cos \tilde{k})^i(1-\cos \tilde{p})^j}{2(1+\epsilon )(1-\cos \tilde{k}%
)+4(1-\cos \tilde{p})}\quad ,  \label{Rij}
\end{equation}
is not transparent enough to provide much insight into shifts in $T_c$. For
the entire range $0\leq \epsilon \leq 1$, we find that $1\leq f<2$, so that
both coefficients in (\ref{TcRatio}) are positive. As a result, $T_c(\alpha
) $ is not a monotonically decreasing function of $\alpha $, as found in
simulations for large $E$. Perhaps we should not be too surprised, since $%
T_c(\alpha ,E)/T_c(1,E)$ can be only a linear combination of $\alpha $ and $%
1/\alpha $. Nevertheless, some features are qualitatively reproduced, such
as $T_c(\alpha _0<1)$ being {\em greater} than $T_c(1/\alpha _0)$. Finally,
noticing that these results are based on an infinite lattice, we extended
them to finite lattices. This generalization consists mainly of replacing
the integrals in (\ref{Rij}) by finite sums over, e.g., $\tilde{k}=2\pi k/L$%
. No significant changes are found down to $L=30$, so that we did not extend
this investigation to $L\neq M$. Again, it is remarkable that qualitative
features, such as the increase of $T_c$ with $L$, are reproduced.

A more serious disadvantage of this approach is that it gives no information
for $E=O(J)$. If the ``high temperature'' expansion is also analytic in $E$,
then the only way $E$ can appear in $T_c(\alpha ,E)$ is through $O(E^2)$.
For reasons beyond the scope of this paper, a second order computation is
prohibitively difficult. As a result, we are unable to extract any insight,
from this approach, into the more interesting parts of the phase diagram.

\section{Conclusions}

In this paper, we report simulation studies of a lattice gas with {\em %
anisotropic} inter-particle interactions, driven to a non-equilibrium steady
state by an external ``electric'' field. Focusing first on two-dimensional
systems ($30\times 30$) subjected to saturation drive, we find the
surprising result that the critical temperature, $T_c(\alpha ,E)$, is a
monotonically {\em decreasing} function of $\alpha \equiv \sqrt{J_{\Vert
}/J_{\bot }}$, for fixed $J\equiv \sqrt{J_{\Vert }J_{\bot }}$ (Fig.1).
Moreover, it is {\em less} than $T_c(\alpha ,0)$ of the equilibrium system
for $\alpha \gtrsim 1.7$! These findings remain essentially unchanged
for systems up to $90\times 90$, leading us to believe that $%
T_c(\alpha ,E)$ for the infinite system is not likely to be drastically
different from those reported here. For drives of the order of $J$, the
transitions from the disordered to an ordered phase are continuous. The
transition temperatures are generally greater than the equilibrium $T_c$'s,
at least within the range of $\alpha \in \left[ 1/3,3\right] $ explored. The
phase diagram is more complex (Fig. 2), since two ordered states (a strip
aligned along or transverse to the drive) appear, depending on $\alpha $ and
$E.$ The transition between H- and V-states is, as expected, discontinuous.
For reasons yet to be fathomed, the line of continuous transitions ``dips''
at the bi-critical point, where it meets the H-V phase boundary. Finally, we
have also investigated {\em rectangular} systems with aspect ratios which
depend on $\alpha $ in such a way that, near criticality, these behave
isotropically with a simple rescaling of the axes. Though some differences
are observed, the phase diagrams are qualitatively unchanged.

On the theoretical front, we have estimated the transition temperatures for
large drives, using the simplest method available: first order in a ``high
temperature'' expansion. Unfortunately, only the most qualitative feature is
reproduced, namely, $T_c(\alpha _0)/T_c(1/\alpha _0)$ is {\em greater} than
unity for $\alpha _0<1$. Since this method is insensitive to $E=O(J)$, the
more interesting aspects of the phase diagram are inaccessible. We hope that
dynamic mean field methods \cite{DMF}, which can probe small drives, will be
successful in producing the behavior in this region of $T$-$\alpha $-$E$
space.

Clearly, this study should be regarded as an initial exploration, providing
us with rough guidelines on which parts of the phase diagram to probe
deeper. Many interesting questions await more detailed simulations as well
as theoretical analyses. We end with a sample list here. The critical
behavior of an equilibrium system is independent of $\alpha $, i.e., whether
the system orders into a V- or an H-state. However, for driven systems,
renormalized field theory predicts drastically different behavior for these
two types of ordering \cite{jslc}. (a) In particular, we expect that
transitions into V-states belong to the universality class of the standard
driven system, with 5 being the upper critical dimension, etc. Thus,
checking the validity of this prediction by simulations would be desirable.
(b) On the other hand, there are no infrared stable fixed points for a
theory with longitudinal phase separation, so that this approach fails to
describe any of the second order transitions into H-states. This kind of
ordering is observed in several other non-equilibrium systems \cite
{basracz,tte}. An interesting question for simulations studies is whether
any of these systems share the same critical behavior, thereby establishing
new classes of driven diffusive systems. (c) Similar issues arise for the
``bi-critical point''. For an equilibrium system, due to the underlying
symmetry, there can be no new behavior. However, for the driven
systems, we expect the two critical lines to fall into different
universality classes, so that there must be non-trivial cross-over phenomena
associated with this point. A similar situation arises in the driven
bi-layer lattice gas \cite{bilayer}. (d) In the context of the above issues,
finite size scaling\cite{fss} stands out as the common and central
analytical tool. (e) For the line of first order transitions at lower $T$,
we could improve the accuracy in several ways. This boundary was determined
by starting runs with a conjectured saddle point configuration (``X'').
Checking with other initial configurations (e.g., ``L'') should increase
confidence in our findings. Running with more sophisticated algorithms,
so as to circumvent the long metastable life-times, would be very desirable.
(f) Finally, the interfaces in the ordered H-states are necessarily
inequivalent, since the drive tends to deposit particles on one and remove
particles from the other. Interfacial correlations in the standard driven
systems (ordered in V-states) display remarkable properties \cite{DDSint}. A
natural question is: what is the nature of the correlation here? In
addition, long wavelength interface instabilities are quite possible \cite
{Mullins,ktl}, so that it will be extremely interesting to investigate
systems with much larger $L$. From this study, it is clear that we are still
far from having an intuitive picture of why driven systems order in the way
they do. Hopefully, with the completion of some of these suggested
investigations, we will be closer to the understanding of these deceptively
simple non-equilibrium systems. 
\begin{eqnarray*}
&& \\
&&
\end{eqnarray*}

\noindent {\Large Acknowledgments\bigskip }

It is a pleasure to dedicate this article to John Cahn, who has devoted much
of his work to the dynamics of diffusive systems within the context of
binary alloys. We are particularly grateful to K. Burns for his initial
efforts at discovering the mysteries of this rich model. We also thank
R. Burghaus for helping us prepare the manuscript.
This research is funded in part by a grant from the National Science
Foundation through the Division of Material Research. One of us (LBS)
acknowledges the generous support by the National Science Foundation through
the Research Experience for Undergraduates program.


\begin{references}
\bibitem{MFTc}  As pointed out in \cite{HTS}, this result can also be
obtained by expanding equation (\ref{Tc(alpha)}) to first order in $J$,
leading to $T_{c}(\alpha )=2\left( J_{\Vert }+J_{\bot }\right) /k_{B}.$
\end{references}
\end{document}